\documentclass[]{spie}  

\usepackage{float}  
 
\usepackage{amsmath,amsfonts,amssymb}
\usepackage{graphicx}
\usepackage[colorlinks=true, allcolors=blue]{hyperref}

\title{MIRAC-5 on the MMT with MAPS: annular groove phase mask N-band coronagraphic upgrade}

\author[a,b]{Alyssa L. Miller}
\author[c]{Jarron Leisenring}
\author[b]{Michael Meyer}
\author[d]{Gilles Orban De Xivry}
\author[d]{Olivier Absil}
\author[b]{Rory Bowens}
\author[d]{Christian Delacroix}
\author[c]{Olivier Durney}
\author[e]{Pontus Forsberg}
\author[c]{Bill Hoffmann}
\author[e]{Mikael Karlsson}
\author[b]{John D. Monnier}
\author[c]{Manny Montoya}
\author[c]{Katie Morzinski}
\author[f]{Eric Pantin}
\author[f]{Samuel Ronayette}
\author[b]{Taylor L. Tobin}
\author[c]{Grant West}

\affil[a]{Applied Physics Program, University of Michigan, Ann Arbor, MI 48103, USA}
\affil[b]{Dept. of Astronomy, University of Michigan, Ann Arbor, MI 48103, USA}
\affil[c]{Steward Observatory and Dept. of Astronomy, University of Arizona, Tucson, AZ 85721, USA}
\affil[d]{STAR Institute, Universit\'{e} de Li\'{e}ge, All\'{e}e du Six Ao\^{u}t 19c, 4000 Li\'{e}ge, Belgium}
\affil[e]{Dept. of Materials Science and Engineering, Angström Laboratory, Uppsala University, Uppsala, Sweden}
\affil[f]{Université Paris-Saclay, Université Paris Cité, CEA, CNRS, AIM, 91191, Gif-sur-Yvette, France}

\authorinfo{Further author information: (Send correspondence to A.L.M.)\\A.L.M: E-mail: alylemil@umich.edu,\\  J.L.: E-mail: jarronl@arizona.edu}

\begin{document} 
\maketitle

\begin{abstract}
We describe the coronagraphic upgrade underway for the Mid-Infrared Array Camera-5 (MIRAC-5) to be used with the 6.5-m MMT telescope utilizing the new MMT Adaptive optics exoPlanet characterization System (MAPS). Mid-IR ground-based coronagraphic adaptive-optics-assisted imaging can be a powerful tool for characterizing exoplanet atmospheres and studying protoplanets in formation within circumstellar disks around young stars. In addition to enabling ground-based observations of bright targets in the background limit, high actuator density 1-2 kHz adaptive optics systems can be competitive with JWST in the contrast limit. We have procured an annular groove phase mask (AGPM) and performed preliminary characterization of its on-axis source rejection as a function of wavelength. We present an optimized Lyot Stop design for use with the AGPM using the High-contrast End-to-End Performance Simulator (HEEPS).  Future work includes implementing the Quadrant Analysis of Coronagraphic Images for Tip-tilt Sensing (QACITS) control loop algorithm with MAPS. We present the system overview, pupil mask design, and expected performance metrics aligned with our scientific goals, building upon recent advances with MIRAC-5 (Bowens et al. 2025) and MAPS. 
\end{abstract}


\section{INTRODUCTION}
\label{sec:intro}  

The technical challenge of high-contrast imaging of exoplanets and circumstellar disks is distinguishing stellar light emanating from the parent star and its surroundings. At visible wavelengths, the flux from cool stellar companions is dominated by the reflected light from the star. However, companions between 300-1000K in effective temperature emit the majority of their energy in the mid-infrared (mid-IR) range\cite{Bowens21, Bowens25}. Therefore, observing companions at thermal infrared wavelengths ($\sim$3-13 $\mu$m) relaxes the stringent contrast requirements set in the reflected-light regime. Thermal emission observations enable better estimations of luminosity, temperature, and atmospheric composition. They also penetrate interstellar dust more effectively than shorter wavelength observations. 

While space-based observations with the James Webb Space Telescope (JWST) have advanced mid-IR exoplanet detection by avoiding atmospheric interference and thermal noise from warm telescope structures\cite{Rigby23}, emerging technologies for ground-based facilities $>$6.5 meters have the potential to complement them.
Ground-based systems in the era of large telescopes supported by adaptive optics (AO) with Strehl $>$90\% \cite{Lei12}, new infrared detectors such as the Teledyne GeoSnap\cite{Lei23}, and coronagraphs could achieve greater contrast-limited performances compared to space-based systems. This has been shown by the Near Earths in the Alpha Centauri Region (NEAR) experiment using the VISIR instrument at the Very Large Telescope (VLT)\cite{Kasper17}, in which the post-processed 3$\sigma$ sensitivity curves reached $\sim 3 \times 10^{-6}$ at 1" for $\alpha$ Centauri A using about 77 hours of observing time in 2019\cite{Wagner21}.

The NEAR project has justified plans for next-generation mid-IR instruments such as the Mid-Infrared ELT Imager and Spectrograph (METIS) on the European Extremely Large Telescope (ELT) to investigate terrestrial planets within the habitable zone\cite{Bowens21}. We are upgrading the Mid-Infrared Array Camera (MIRAC-5) system, which is coupled with the MMT AO exoPlanet characterization System (MAPS)\cite{Morzinksi20} at the 6.5-m MMT Telescope, by implementing a coronagraph. This upgrade will significantly improve MIRAC-5's high-contrast imaging capabilities, enabling more sensitive observations\cite{Rigby23, Bowens25, Bowens22}.
 
In this paper, we first describe the MMT/MIRAC-5 system with MAPS AO in Section~\ref{sec:System Overview} before introducing the coronagraph design and predicted performance in Section~\ref{sec:Coronagraph}. In Section~\ref{sec: Schedule}, we close with our projected schedule for this coming fall and our plans for precision control to enable optimal coronagraphic performance at the MMT.

\section{System Overview}
\label{sec:System Overview}

\subsection{MMT and MAPS AO}
\begin{figure}[H]
   \begin{center}
   \begin{tabular}{c} 
   \includegraphics[height=7.5cm, width = 14cm]{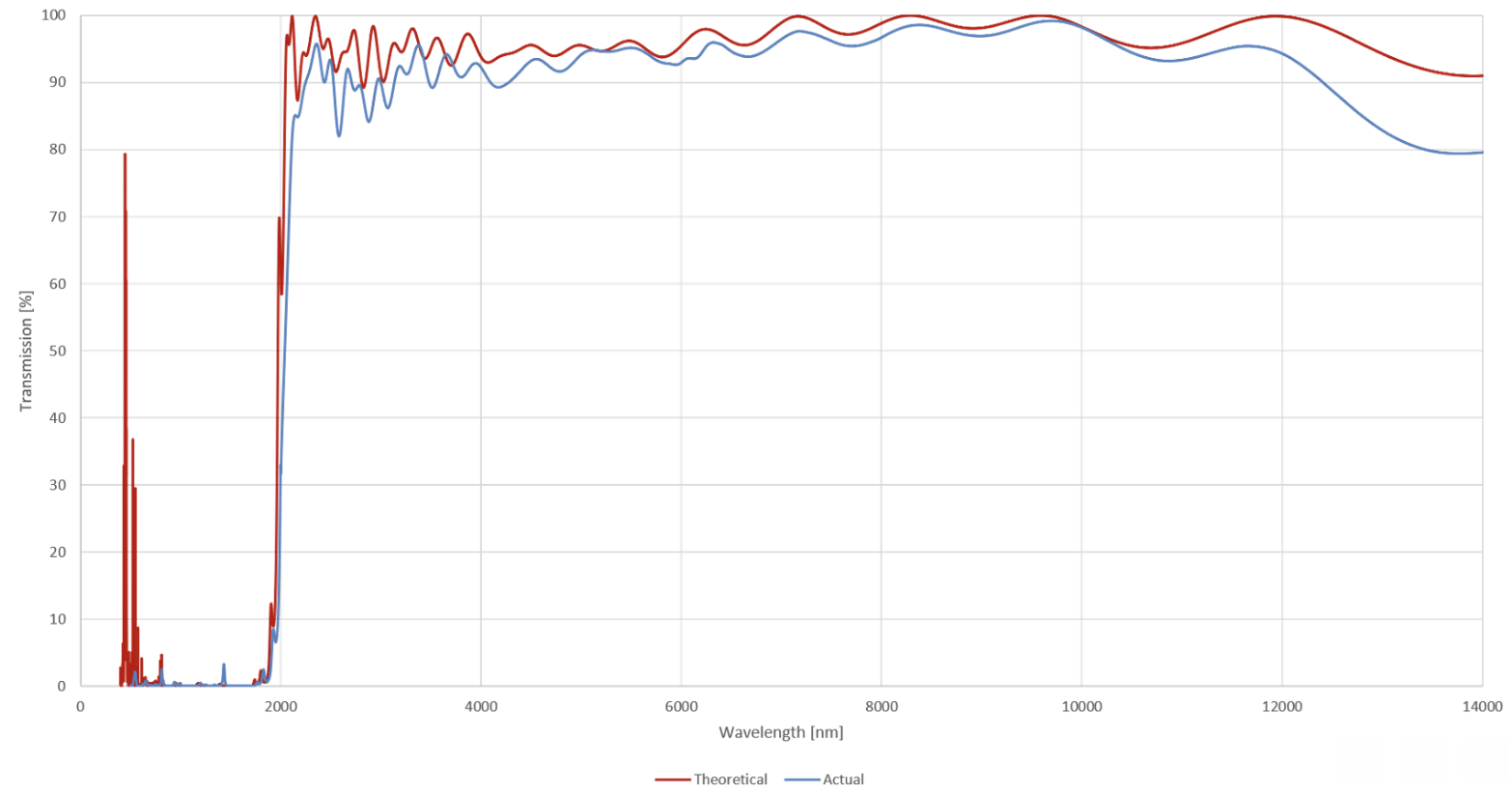}
   \end{tabular}
   \end{center}
   \caption[Dichroic] 
   { \label{fig:dichroic} 
The transmission from 2-15 $\mu$m of the ZnSe dichroic that is used to split the light between MAPS and MIRAC-5. Light is reflected below 2 $\mu$m to MAPS for wavefront sensing.}
   \end{figure}

\begin{figure}[H]
   \begin{center}
   \begin{tabular}{c} 
   \includegraphics[height=8.5cm, width = 11cm]{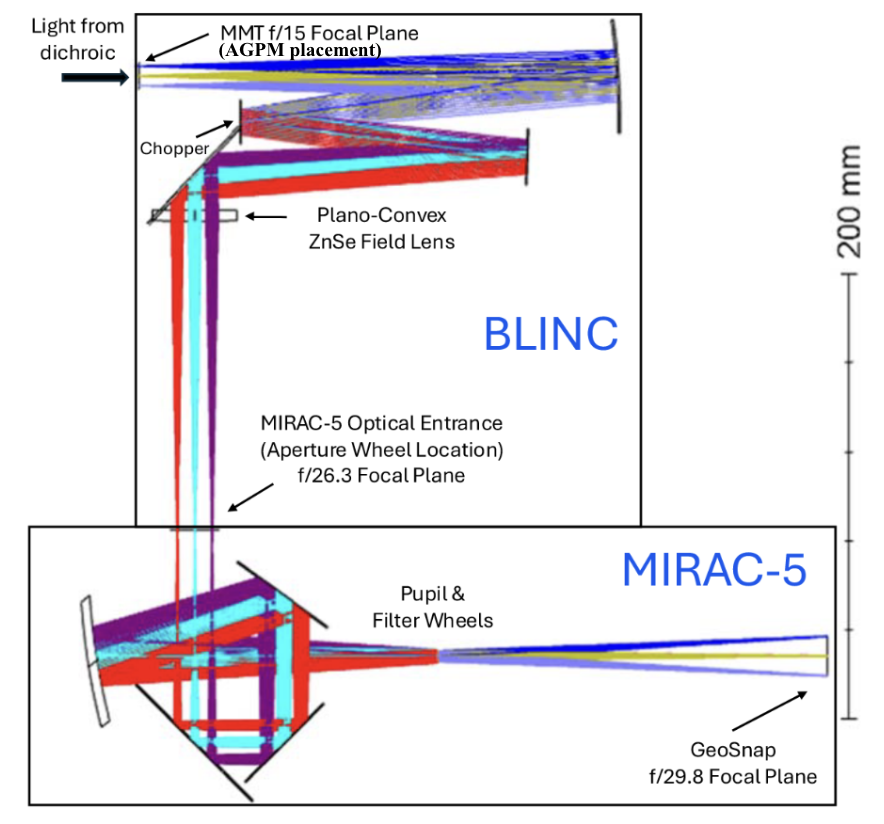}
   \end{tabular}
   \end{center}
   \caption[System] 
   { \label{fig:System} 
Current BLINC and MIRAC-5 optical layout as presented in Bowens et al.(2022, 2025) with the AGPM located at the f/15 focal plane.}
   \end{figure}

The MMT Telescope features a monolithic, 6.5-meter primary mirror with a f/15 adaptive secondary mirror (ASM). MAPS AO is currently undergoing commissioning at the MMT. This adaptive optics system uses 336 voice coil actuators operating at 1-2 kHz to correct 100-220 modes with 1-ms delay response time, using two high performance visible and near-infrared (NIR) pyramid wavefront sensors (WFS)\cite{Morzinksi20}. 
Following the ASM, the telescope light will pass through a ZnSe dichroic, which is used to split the light between the MAPS instrument assembly and MIRAC-5. The dichroic transmits wavelengths (as shown in Figure \ref{fig:dichroic}) from 2-15 $\mu$m with broadband transmission $>$95\%, and reflects wavelengths $<$2$\mu$m to MAPS for wavefront sensing using either the optical or NIR cameras. The beam then enters the BLINC liquid nitrogen cooled cryostat, which contains the pupil plane chopping mechanism, and is used to reimage the Cassegrain focus on the detector within the MIRAC-5 cyrostat (which is cooled with a closed-looped He cyrocooler) as shown in Figure \ref{fig:System}. 

\subsection{MIRAC-5 Detector and Capabilites} 
The MIRAC-5 camera is a HgCdTe (MCT) $1024 \times 1024$ pixel array detector that is sensitive within its half-peak QE cut-offs between 2 to 13 $\mu$m.\cite{Bowens25, Lei23} This detector is well-suited to the high-backgrounds of ground-based systems for mid-IR observations with its 85 Hz framerate, 1.2 million electron well-depth, and 18 $\mu$m pixel pitch.  MIRAC-5 provides $\sim$ 19 arcseconds of field of view and has a pixel scale of 0.0192 arcsec/pixel\cite{Bowens25}. With MAPS AO, we anticipate $>$90\% Strehl for near-diffraction-limited N2-band (10-12.5 $\mu$m) observations. 

The annular groove phase mask (AGPM) will be mounted at the f/15 focal plane of the MMT as shown in Figure \ref{fig:System}) prior to the chopping mechanism and cooled to $\sim$77K. This differs from what was done in the NEAR system as this will allow the source to modulate between the two chopping positions on the MIRAC-5 detector while remaining suppressed by the coronagraph. With the addition of the AGPM, MIRAC-5 on the MMT paired with MAPS AO will improve contrast-limited performance to $\sim \lambda/D$ separations ($\sim0.3$"). The MIRAC-5 AGPM will be complemented by a cold stop pupil mask (Lyot Stop) in the relayed pupil plane where the filter and pupil wheels are indicated in Figure \ref{fig:System} to suppress diffracted stellar light.

\section{Coronagraphic Implementation and Optimization}
\label{sec:Coronagraph}
 
\subsection{Annular Groove Phase Mask (AGPM)}

The AGPM is a type of vortex coronagraph consisting of concentric, subwavelength gratings placed at the focal plane. This vortex phase mask induces a vortex phase shift that suppresses on-axis stellar light by redirecting it from the on-axis source to the outside of the downstream geometric pupil image.\cite{Mawet05,Absil16}

\begin{figure}[H]
   \begin{center}
   \hspace{-1.3cm}
   \includegraphics[height=6.8cm, width=11cm]{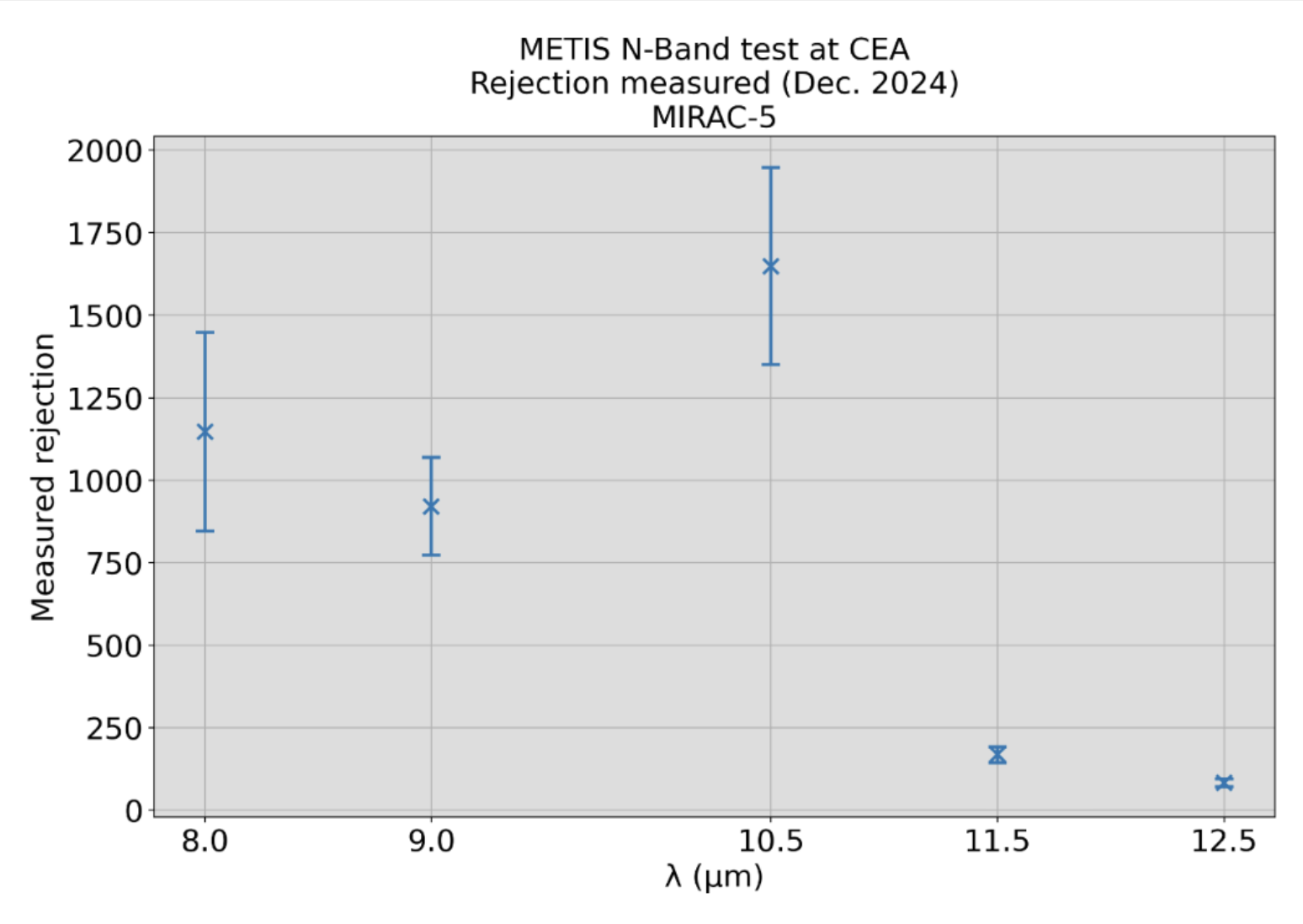}
   \hspace{0.3cm} 
   \raisebox{3.2cm}{%
       \begin{minipage}{4cm}
          \centering
\footnotesize 
\begin{tabular}{|c|c|}
\hline
\parbox[c][1cm][c]{2cm}{\centering $\lambda$ ($\mu$m)} & \parbox[c][1cm][c]{2cm}{\centering Rejection \\ Factor} \\
\hline
\parbox[c][0.5cm][c]{0.5cm}{\centering 8.0} & \parbox[c][0.5cm][c]{2cm}{\centering 1148 $\pm$ 300} \\
\hline
\parbox[c][0.5cm][c]{0.5cm}{\centering 9.0} & \parbox[c][0.5cm][c]{2cm}{\centering 920 $\pm$ 150} \\
\hline
\parbox[c][0.5cm][c]{0.5cm}{\centering 10.5} & \parbox[c][0.5cm][c]{2cm}{\centering 1650 $\pm$ 300} \\
\hline
\parbox[c][0.5cm][c]{0.5cm}{\centering 11.5} & \parbox[c][0.5cm][c]{2cm}{\centering 168 $\pm$ 25} \\
\hline
\parbox[c][0.5cm][c]{0.5cm}{\centering 12.5} & \parbox[c][0.5cm][c]{2cm}{\centering 83 $\pm$ 12} \\
\hline
\end{tabular}
       \end{minipage}%
   }
   \end{center}
   \caption[example] 
   { \label{fig:measure} The MIRAC-5 AGPM component on-axis starlight rejection measured with a laser versus wavelengths at CEA Saclay in France.
   }
\end{figure}

We evaluated an available N2 (10-12.5 $\mu$m) optimized AGPM that was manufactured following the success of the METIS AGPM manufacturing\cite{Delacroix24}. The AGPM's rejection performance for the on-axis stellar light was measured using a cryogenic testbed at CEA Saclay as shown in Figure \ref{fig:measure}. This testbed consists of a telescope simulator that is fed with five IR quantum cascade lasers that are used as point-like sources to illuminate the AGPM as described in Delacroix et al. (2024)\cite{Delacroix24}. This AGPM has achieved a rejection factor of about $\leq 1600$:1 at 10.5 $\mu m $. Based on this performance, we adopted this AGPM (which is 10 mm in diameter) for implementation in the MIRAC-5 system. 


\subsection{Lyot Stop Design}
The vortex phase mask can theoretically provide total on-axis stellar light rejection for a circular, unobstructed pupil.\cite{Absil16, Shinde21}Due to the presence of the central obscuration, spider support vanes, and phase errors, this theoretical performance breaks down, leading to stellar leakage \cite{Shinde21}. The Lyot Stop is used to block the diffracted stellar light outside of the pupil by the AGPM. It can be adapted to mitigate most stellar leakage by undersizing the outer diameter and oversizing the inner diameter and spider widths associated with the entrance pupil. The MIRAC-5 AGPM Lyot Stop should balance both contrast and throughput to maximize the signal-to-noise (SNR) of a faint companion.

For the purpose of obtaining simulated high-contrast performance, a series of $\sim$1-hour long AO contrast sequences were produced using the High-Contrast End-to-End Perforance Simulator (HEEPS)\cite{Carlomagno20, Delacroix22,Shinde21} python package, which relies on the PROPER optical propagation library to produce noiseless coronagraphic point-spread-functions (PSFs)\cite{Krist07}. HEEPS also includes the Vortex Imaging Processing (VIP) package\cite{Val23} to produce simple mock post-ADI processing sensitivity curves.
The simulated optical layout only includes the pupil planes and focal planes. The coronagraph, the Lyot Stop, and the detector are placed in the first focal plane, pupil image plane, and the second focal plane, respectively.

To perform realistic high-contrast imaging (HCI) performance estimates, the HEEPS pipeline involves using a temporal series of AO residual phase screens from an end-to-end AO simulation tool (COMPASS)\cite{Gratadour14} to propagate these screens through the HCI elements using PROPER\cite{Krist07}. We used the GRAAL COMPASS AO residual phase screens that were created for the NEAR experiment at the VLT.\cite{Maire20} The wavefront error of these phase screens is approximately 100-200 nm rms for the simulations. This AO residuals cube was used to generate realistic, but optimistic results as the wavefront error for MAPS is projected to be approximately 200-300 nm rms\cite{Morzinksi20} as opposed to GRAAL, which is an adaptive optics module for the VLT\cite{Jer16}. These simulated phase screens are $\sim$1 hour sequences sampled every 100 ms. Due to the simulated loop closure, the wavefront error had not yet converged, so the first $\sim$200 screens ($\sim$2 seconds) were excluded from the analysis. To save computational time, the original 36,000 frames of the AO sequence were decimated down to 3600 frames (sampling every 10 frames) while maintaining the temporal integrity of the sequence. This corresponds to one AO phase screen to one single detector integration time (i.e DIT = 1 s). 

\begin{table}[H]
\caption{Main parameters used for the end-to-end propagation for the purpose of optimizing the Lyot Stop mask. } 
\label{tab:configs}
\begin{center}       
\begin{tabular}{{l@{\hspace{2cm}}l}} 
 
\hline\hline 
\rule[-1ex]{0pt}{2.5ex}  HEEPS Configuration & Value  \\
\hline 
\rule[-1ex]{0pt}{2.5ex}  Pupil diameter (eff. diam ext.) & 6.37 meters \\
\rule[-1ex]{0pt}{2.5ex}  Nominal diam (for mask oversizing) & 6.37 meters  \\
\rule[-1ex]{0pt}{2.5ex}  Pupil Image Size (for PROPER) & 6.7 meters  \\
\rule[-1ex]{0pt}{2.5ex} Central Obscruation & 0.9 meters  \\
\rule[-1ex]{0pt}{2.5ex}  Spider width (approx.) & 0.1 meters \\
\rule[-1ex]{0pt}{2.5ex} Segmentation & None \\
\rule[-1ex]{0pt}{2.5ex}  Latitude (MMT) & 31.6889 deg  \\
\rule[-1ex]{0pt}{2.5ex} Apparent $M_{N_{2}}$ (Procyon)  & -0.63 \\
\rule[-1ex]{0pt}{2.5ex} Declination & 5.22 deg\\
\rule[-1ex]{0pt}{2.5ex}  Detector Angular Pixel Scale & 0.038 arcsec \\
\rule[-1ex]{0pt}{2.5ex}  Half Field of View  (rel. to 19" FOV) & 6 arcsecs   \\
\rule[-1ex]{0pt}{2.5ex} Detector Integration Time (DIT) & 1 sec  \\
\rule[-1ex]{0pt}{2.5ex}  Grid Size& 2048 $\times$ 2048 \\
\hline 
\end{tabular}
\end{center}
\end{table}

The entrance pupil used for these simulations is an annular pupil with spiders of the effective external diameter of the MMT, which is defined by the ASM. The ASM was designed to be undersized for the primary mirror to reduce thermal background emission coming from the primary. So, the effective external diameter is 6.37 meters. Considering the central obscuration, the primary mirror's central hole is 0.898 m ($\sim 14.1$\%) of effective external diameter, which does encompass an additional central obscuration in the ASM. The parameters in Table \ref{tab:configs} were used to generate the mock contrast curves. We modified the pixel scale and grid size to achieve finer pupil sampling ($\sim$225 pixels across the pupil diameter) while avoiding computationally expensive large matrices (e.g., $4000 \times 4000$ grid size). Note that this pixel scale differs from the actual system value given in Section \ref{sec:System Overview}.

\begin{figure}[H]
    \hspace{-3cm}
    \centering
    \begin{minipage}[c]{0.45\textwidth}
        \centering
        \includegraphics[width=5.0cm]{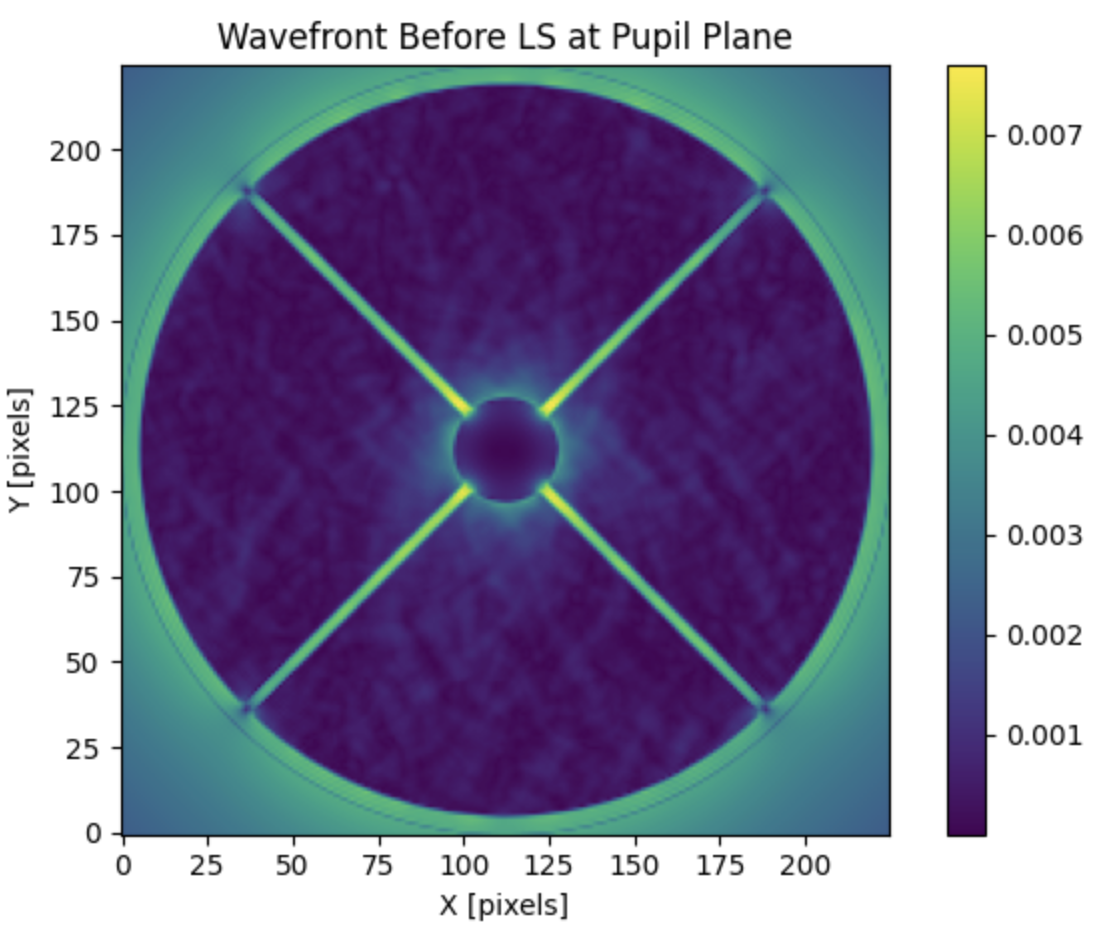}
        \vspace{0.2cm}
        
        \includegraphics[width=5.2cm]{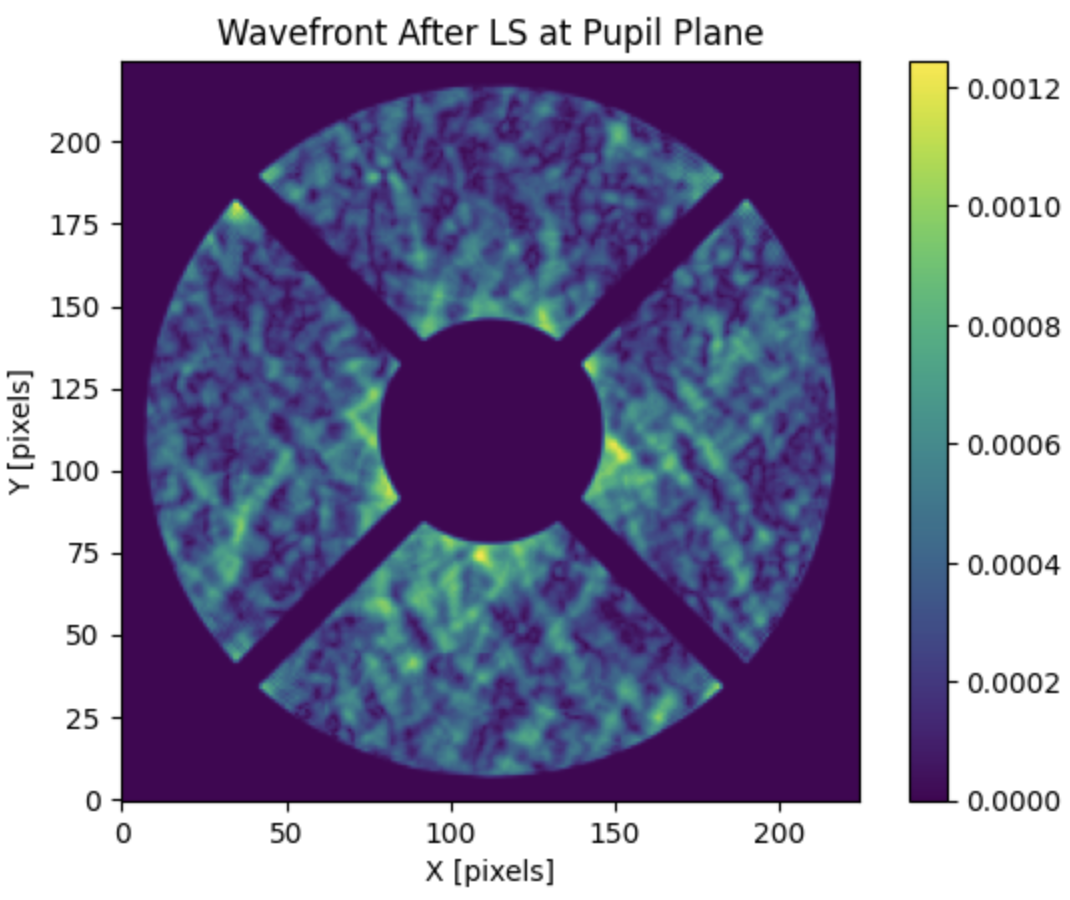}
    \end{minipage}
    \hspace{-0.4cm}
    \begin{minipage}[c]{0.45\textwidth}
        
        \centering
        \includegraphics[width=10cm, height=7cm]{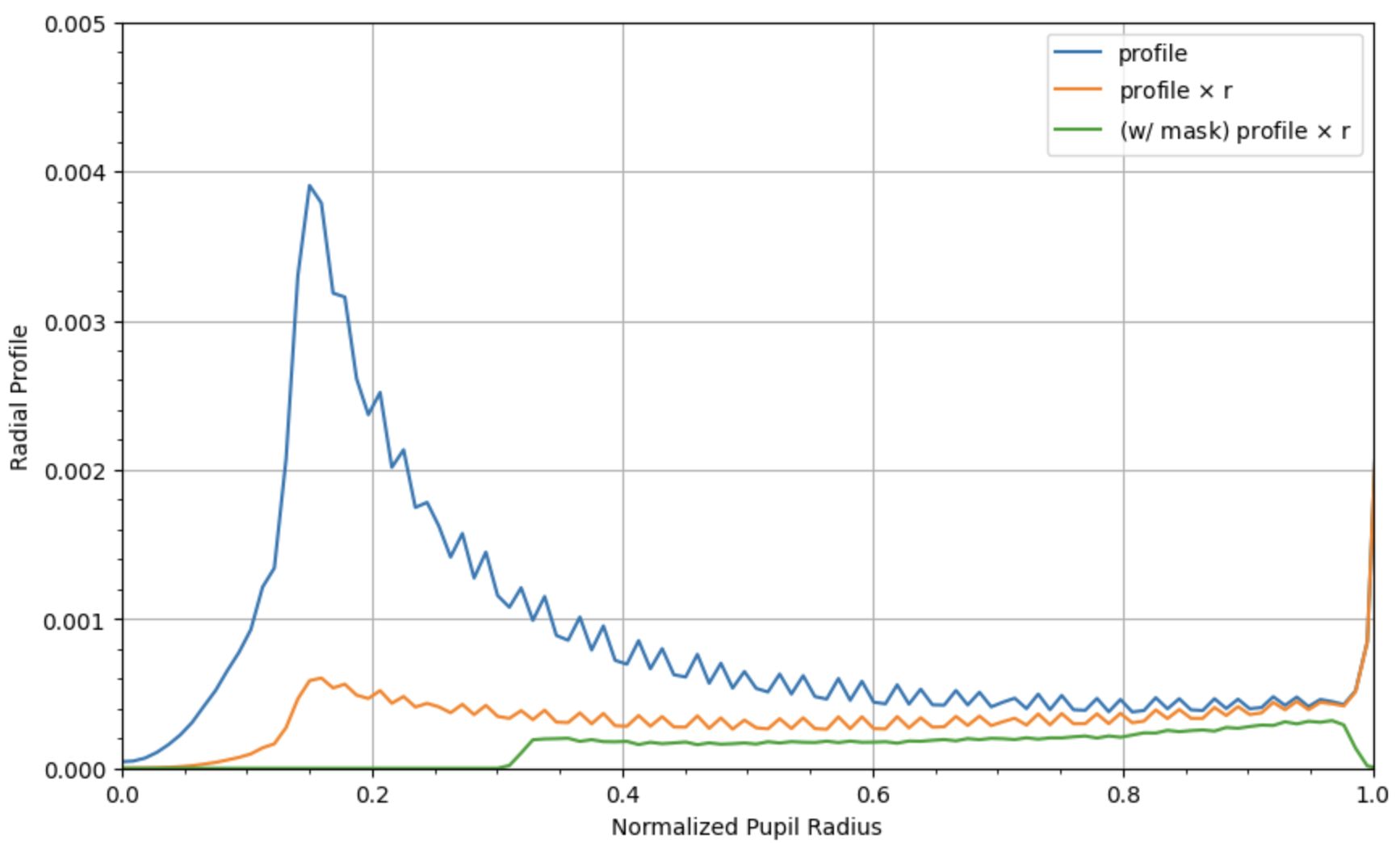}
    \end{minipage}
    
    \caption{Left: the intensity distribution in the pupil plane at N2 after the AGPM and just before the Lyot Stop (top) and after the Lyot Stop (bottom). Right: the corresponding intensity profile  showcasing the stellar leakage (blue) and the profile multiplied by the radial distance from the center of the pupil without the mask (orange) and with the mask (green). }
    \label{fig:pupils}
\end{figure}

The HEEPS algorithm implemented a classical vortex coronagraph (CVC) mode, which represents using an AGPM, in the N2 filter for the MIRAC-5 system at the MMT. Because at least 30 degrees of field rotation is required for optimal post-processing, the target star, Procyon (Dec. +5.2 degrees where the lattiude of MMT is +30), was chosen to be representative of the coronagraph capabilities with MIRAC-5 for bright stars in about 1 hour of observation. The MIRAC-5 SNR Estimator\cite{Bowens25} was used to obtain the stellar flux and the instrumental background given in Table \ref{tab:Modes}.
Assuming an ideal coronagraph alignment, Figure \ref{fig:pupils} showcases the re-imaged pupils generated through HEEPS, which represents the normalized amplitudes of the electric field derived from the AO residual wavefront at the downstream pupil plane. The purpose is to visualize the simulated residual on-axis light suppression made by the AGPM and the Lyot Stop in the system.

\begin{table}[H]
\caption{Selected mode, bandwidth, and additional parameters from MIRAC-5 SNR Estimator} 
\label{tab:Modes}
\begin{center}    
\begin{tabular}{|c|c|c|c|c|} 
\hline
\renewcommand{\arraystretch}{3.0} 
\rule[-1ex]{0pt}{3.5ex} \parbox{3cm}{\centering Coronagraphic \\ Mode}&  Band & \parbox{3cm}{\centering Simulated \\ Wavelength ($\mu$m)} & \parbox{3cm}{\centering Stellar Flux \\ (e-/s)} & \parbox{3cm}{\centering Instrumental \\ Background \\ (e-/s/pix)} \\ 
\hline
\renewcommand{\arraystretch}{2.0} 
\rule[-1ex]{0pt}{3.5ex} {\centering CVC} & {\centering N2} & {\centering 11.20} & {\centering 9.307e+08} & {\centering 1.50e+07}\\
\hline 
\end{tabular}
\end{center}
\end{table}
The stellar flux and instrumental background values were obtained with the framerate set to 85 Hz (0.012 seconds) with the MIRAC-5's well-depth of 1.2 million electrons\cite{Bowens25}.

\subsection{Predicted Contrast Performance}

\begin{figure} [H]
   \begin{center}
   \begin{tabular}{c} 
   \includegraphics[height=8cm, width = 15cm]{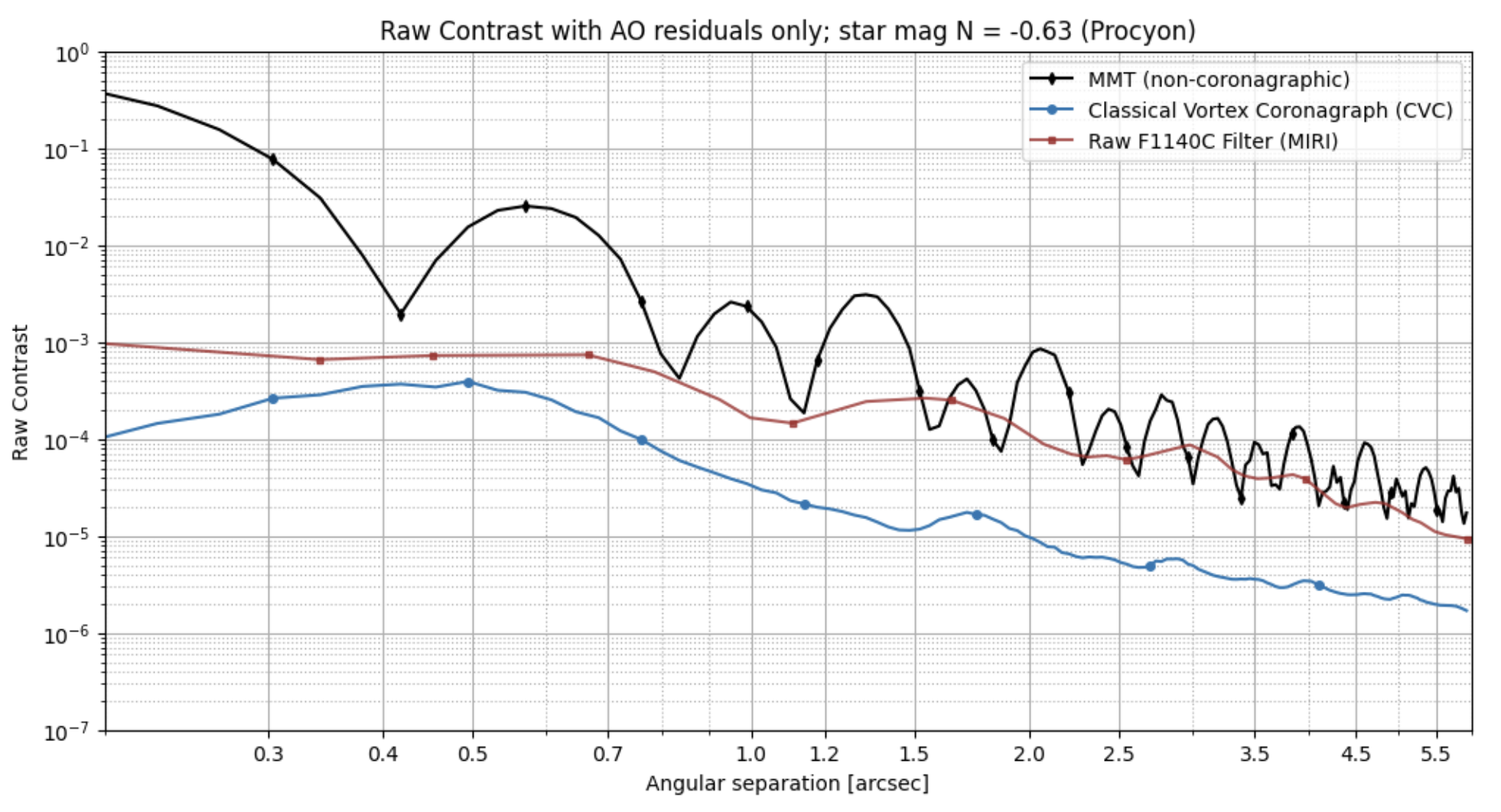}
   \end{tabular}
   \end{center}
   \caption[example] 
   { \label{fig:Raw} The azimuthally average point-spread-function (PSF) normalized peak between the off-axis source transmission and the target star for MMT without the coronagraph (black), as well as using the classical vortex coronagraph (CVC) mode which represents the use of the AGPM coronagraph (blue). Included is the raw contrast of the MIRI data using the four-quadrant phase mask\cite{Rou00} for the F1140C filter (brown) from Figure 5 from Boccaletti et al (2022)\cite{Bocc22}.}
   \end{figure}

The simulations through HEEPS produce noiseless, speckle-limited coronagraphic PSFs for each residual AO screen and were performed at a single-wavelength value, but are considered representative for N2 observations between 10-12.5 $\mu$m. The raw contrast curves shown in Figure \ref{fig:Raw} represent the radial profile of the azimuthally averaged intensity of the simulated PSFs of the on-axis source, normalized by the peak intensity of the off-axis PSF radial profile.
Incorporating the AGPM coronagraph with ideal alignment in MIRAC-5 with MAPS AO will be able to achieve a raw contrast on the order of $\sim 3 \times10^{-5}$ around 1". The MIRAC-5 AGPM raw contrast is projected to be better at 1" compared to the Mid-Infrared Instrument (MIRI) F1140C filter (11.40 $\mu$m, $\Delta\lambda = 0.8$ $\mu$m), which has a raw contrast of $\sim2\times10^{-4}$ at 1" on the JWST according to Figure 5 from Boccaletti et al. (2022). \cite{Rigby23, Bocc15, Bowens22} The MIRI raw contrast data shown in Figure \ref{fig:Raw}, obtained using the four-quadrant phase mask (4QPM) coronagraph,\cite{Rou00,Bocc22} were digitally extracted using WebPlotDigitizer\cite{WebPlotDigitizer}.

\begin{figure} [H]
   \begin{center}
   \begin{tabular}{c} 
   \includegraphics[height=8cm, width = 15cm]{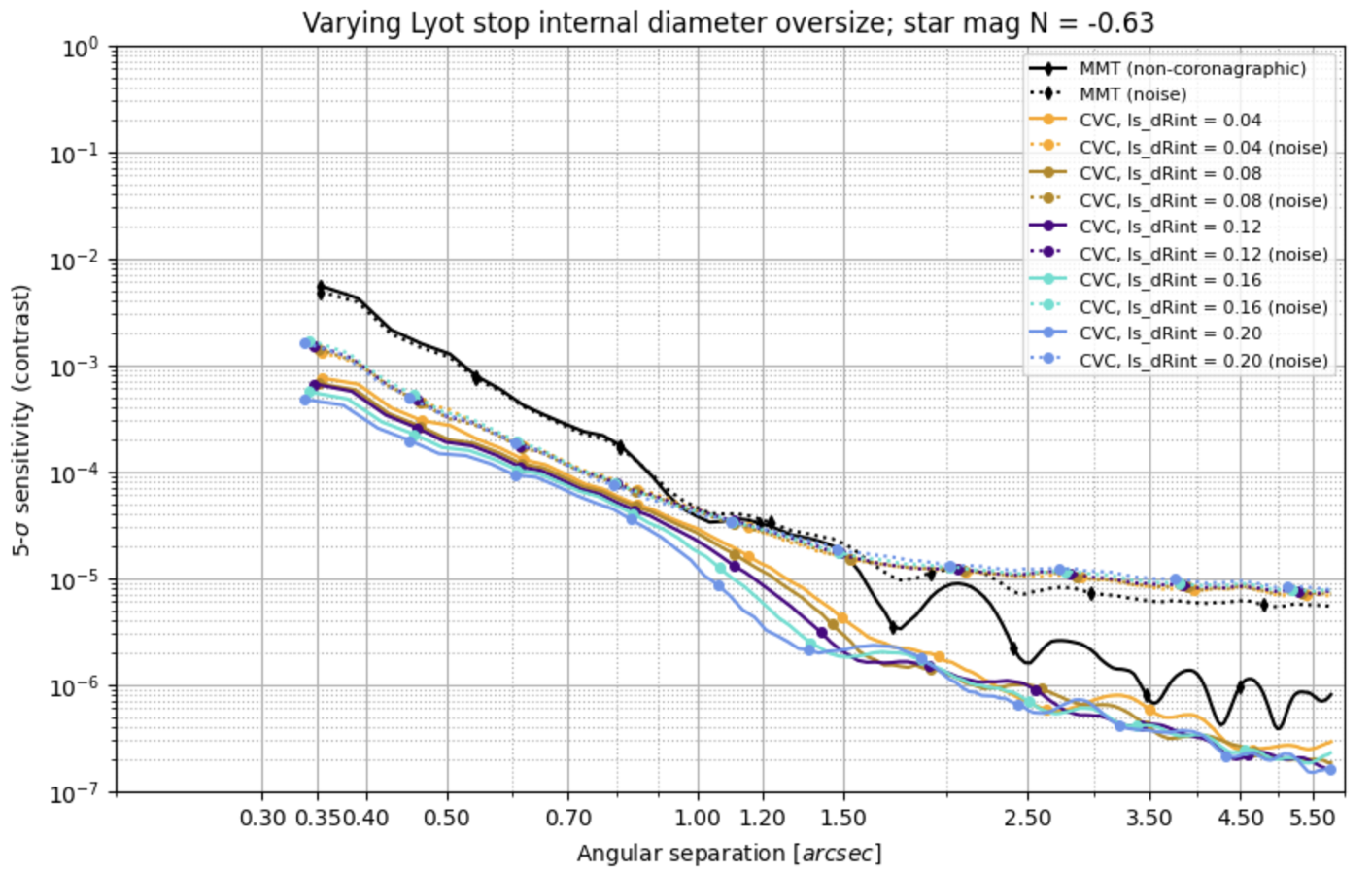}
   \end{tabular}
   \end{center}
   \caption[example] 
   { \label{fig:ADI} Post-processing 5$\sigma$ sensitivity curves with varying inner radius of the Lyot Stop in the speckle-limited case (solid lines) compared to speckle $+$ source $+$ background-noise (dotted lines). The ls\_dRint variable represents the inner diameter percent of the effective external diameter of the entrance pupil. The current mask is projected to fall in between the solid turquoise and light blue lines above (0.16 and 0.20 respectively).}
   \end{figure}

To understand the detection threshold for companions at 5$\sigma$ sensitivity after subtracting the stellar PSF, the on-axis PSF cube is processed using the classical median-subtraction ADI technique using the VIP package through HEEPS. The 5$\sigma$ contrast limits are computed by first injecting mock planetary companion signals into the data prior to this processing technique to model the effect of the algorithm on point sources using VIP's built-in contrast calculation\cite{Delacroix22, Val23}. In addition, the background and photon-noise that are added are represented by the dotted post-ADI curves seen in Figure \ref{fig:ADI}. The solid lines in Figure \ref{fig:ADI} represent the speckle-limited performance without background added.

The central obscuration (as shown in Figure \ref{fig:pupils}) exhibits the greatest stellar leakage that needs to be blocked off by the Lyot Stop's inner diameter. We varied the inner radius of the Lyot Stop, and set the Lyot Stop external diameter to be undersized at 98\% of the effective diameter of the MMT pupil in N2 and the spider masks oversized to $\sim$3.2\% of the effective diameter to account for pupil stability and uncertainty. From these simulations, we see minimal gain in contrast between increases of the inner diameter inside 1". 

The simulated sequences of Procyon are background-limited beyond 1.5". With our focus between 0.35" - 0.90" in the contrast-limited regime, and with minimal differences between the system throughput and contrast-gain, the current pupil mask already installed in MIRAC-5 can be a suitable candidate for the MIRAC-5 AGPM. 

The current pupil mask is made of phosphor bronze coated with black Aeroglaze Z306, which adds approximately 30-50 $\mu$m of thickness to the component. This mask was designed to have an inner diameter of 18\% of the effective external diameter (oversized by a total of 28\% from the central obscuration shown in Table \ref{tab:configs}), which falls between the two post-ADI sensitivity curves for the CVC mode of a fraction of the effective diameter of 0.16 and 0.20 as shown in Figure \ref{fig:ADI}. The throughput of our simple model for the MIRAC-5 coronagraph through HEEPS projects a total off-axis source transmission of $\sim$ 74.8\%. 

Based on Figure \ref{fig:ADI}, we are projected to reach a 5$\sigma$ post-processed sensitivity using classical ADI, reaching speckle-noise limit, of $\sim1.5 \times 10^{-5}$ at about 1". However, more sophisticated post-processing could improve this by as much as a factor of 10\cite{Wagner21}. To reach this sensitivity level (in this case for a bright star like Procyon) the approximate observing time will be a few hours. Boccaletti et al.(2022) reported that PCA post-processing for MIRI on JWST could achieve $\sim2 \times 10^{-5}$ inside 1" at 3$\sigma$. As this appears to be background-limited, observations of brighter stars are needed to better assess actual performance. MIRAC-5's current pupil mask and the new AGPM should be competitive with the MIRI coronagraphs in N2 observations in the contrast-limited regime and have high-throughput for off-axis companions.

With the addition of the AGPM, MIRAC-5 should be capable of characterizing high priority targets 
between 1-2 $\lambda$/D, competitive with JWST MIRI. For example, based on the Sonora Bobcat model\cite{Mar21}, 51 Eri b\cite{Sam17} is expected to have a contrast of $\sim6 \times 10^{-4}$ at 0.46" projected separation. Another feasible target is Kappa And b with a comparable contrast requirement in the N2 filter of $\sim4 \times 10^{-4}$ at 0.93" projected separation.  This is, in principle, feasible on the basis of the
contrast curves shown in Figure \ref{fig:ADI}.  However, we note that 51 Eri b is a much fainter host compared to Procyon. It would take nearly an entire night of observation to reach the sensitivity where we 
are contrast-limited at this level for a star this faint (N2 $\sim$ $5^{th}$mag) as discussed in Bowens et al. (2025).  

\section{Future Work}
\label{sec: Schedule}

During the Fall of 2025, the AGPM focal plane mask will be installed and we will verify the alignment between MIRAC's internal Lyot Pupil Stop and the telescope pupil. Following the installation, we will measure the first on-sky throughput and contrast performance across the field of view with MIRAC-5 and updated MAPS AO performance. The new dichroic was installed in late 2024. We plan to measure the system's throughput with this dichroic, as this will be only its second deployment on the telescope. The ambient temperature conditions (and thus telescope background) are also expected to differ from its first use\cite{Bowens25}. While we obtain the first on-sky measurements with the AGPM, we will test the various observing modes with MAPS AO and characterize the PSFs and contrasts. We hope to demonstrate speckle-limited raw contrast around bright stars in Fall 2025, and 
long-exposure speckle-limited observations in Spring 2026. 

With the small inner working angle of the AGPM\cite{Mawet05}, the system is sensitive to pointing drifts and will require a precise control system to allow the coronagraph to perform optimally for long exposures. Future work will consist of validating the current operations model for the Quadrant Analysis of Coronagraphic Imaging for Tip-Tilting (QACITS) control loop algorithm\cite{Huby17, Maire20} with MAPS AO. This control system will be calibrated with respect to our system in order to keep target stars centered on the AGPM during pointing offsets. We hope to prove that the MIRAC-5/MAPS  system with a coronagraph will obtain sensitivities similar to the NEAR experiment, and be competitive with the JWST/MIRI coronagraphs in the contrast-limited regime using sophisticated post-processing.

\begin{center}
{\large \textbf{Acknowledgments}}
\end{center}
This project was funded by the Heising-Simons Foundation through grant 2020-1699.

\bibliography{report} 
\bibliographystyle{spiebib} 

\end{document}